\documentclass[aps, prd,twocolumn, showpacs, showkeys,nofootinbib,floatfix]{revtex4-1}

\usepackage{amsmath}
\usepackage{amsfonts}
\usepackage{txfonts}
\usepackage{graphicx}
\usepackage{dcolumn}
\usepackage{bbm}
\usepackage{amssymb}
\usepackage{latexsym}
\usepackage{CJK}
\usepackage{titlesec}
\usepackage[colorlinks=true, linkcolor=red, citecolor=blue]{hyperref}

\begin{document}

\title{Confronting DGP Braneworld Gravity with Cosmic Observations after {\it Planck} Data}
\author{Lixin Xu}
\email{lxxu@dlut.edu.cn}
\affiliation{Institute of Theoretical Physics, School of Physics \&
Optoelectronic Technology, Dalian University of Technology, Dalian,
116024, P. R. China}

\date{\today}

\begin{abstract}
The normal branch of Dvali-Gabadadze-Porrati braneworld gravity with brane tension is confronted by the currently available cosmic observations from the geometrical and dynamical perspectives. On the geometrical side, the type Ia supernova as standard candle, the baryon acoustic oscillation as standard ruler and the cosmic microwave background measurement from the first released 15.5 months data were used to fix the background evolutions. On the dynamical side, the redshift space distortion data will be used to determine the evolution of the matter perturbation. Through a Markov chain Monte Carlo analysis, we found the dimensionless crossover scale $\Omega_{r_c}=1/(4H^2_0r^2_{c})=0.00183_{-0.00183}^{+0.000338}$  in a spatially flat normal branch of Dvali-Gabadadze-Porrati braneworld. This result suggests that the crossover scale $r_c$ should be around $12H^{-1}_0$ which is consistent with the previous result $r_c>3H^{-1}_0$ and greater. It also implies that the five-dimensional gravity effect is weak to be observed in $H^{-1}_0$ scale.
\end{abstract}


\maketitle

\section{Introduction}

Based on the high-dimensional gravity theory, Dvali, Gabadadze and Porrati (DGP) proposed a braneworld model \cite{ref:DGP} in which our Universe is a four-dimensional brane (a hypersurface) embedded in a five-dimensional Minkowski space-time, please see Ref. \cite{ref:DGPreview} for a review. In this braneworld scenario, the gauge forces are confined on the brane, but the gravity can propagate in all dimensions freely. An infrared modification to gravity is achieved due to the existence of the so-called crossover scale $r_c$. Below the crossover scale $r_c$, the gravity appears four-dimensional. However, above the scale $r_c$ the gravity can leak into the extra dimension and make the conventional four-dimensional gravity altered. It is due to the leakage of gravity, the current Universe appears an accelerated expansion phase \cite{ref:Riess98,ref:Perlmuter99}. There are tow branches of cosmological solutions in the DGP model, named self-accelerating branch and normal branch. In the self-accelerated branch, late-time accelerated expansion of our Universe occurs without the need of a cosmological constant. But in this case, ghost degrees of freedom in the linearized theory \cite{ref:DGPghost} and substantial conflict with cosmic data \cite{ref:DGPdataconflict,ref:DGPxu} appear. Although, a cosmological constant or brane tension can alleviate the conflict with cosmic observation, nonlinear solutions are required to solve the ghost problem \cite{ref:DGPghostsol}. On the normal branch, the late-time accelerated expansion can be realized after including a cosmological constant. In this case, no ghost appears. Therefore, in this paper, we will mainly focus on the normal branch with the aid of a cosmological constant.          

To test the gravity on large scales, the only possible tool is the cosmic observations which include the geometrical and dynamical measurements. Through the geometrical measurements, for example the type Ia supernova (SN), the baryon acoustic oscillation (BAO) and the cosmic microwave background (CMB), the background evolution history can be fixed. When the information of the large scale structures is included, the dynamical evolution, such as the growth of the matter perturbation, can be determined. Combining the two sides measurements, the degeneracies between model parameters would be broken; and a tight constraint can be obtained. In this paper, we conduct a Markov chain Monte Carlo (MCMC) study of the normal branch of the DGP model by using the currently available cosmic observations which include SN, BAO and first released 15.5 months CMB data from {\it Planck} in the geometrical side and redshift space distortion (RSD) data in the dynamical side.         

Actually, the DGP braneworld cosmology has been constrained by various cosmic observations in the past years with and without a cosmological constant in the two branches \cite{ref:DGPdataconflict,ref:DGPxu,ref:DGPconstraint1, ref:DGPconstraint2,ref:DGPconstraint3,ref:DGPPPF}. Importantly, Lombriser {\it et al} \cite{ref:DGPPPF} constrained the two branches of the DGP braneworld cosmology exhaustively by adopting a parameterized post-Friedmann (PPF) description of gravity, where all of the information from the CMB data, including the large scales, and its correlation with galaxies in addition to the geometrical constraints from supernovae distances and the Hubble constant were utilized. Where the requirement of a tension or a cosmological constant was confirmed; And they concluded that the crossover scale must be greater than the Hubble scale $H_0r_c > 3$ and $3.5$ at the $95\%$ C.L. with and without a spatial curvature. It was also pointed out the importance of using the large scale structure information to constrain the infrared modifications to gravity. Therefore, in this paper, we try to confront the normal branch with a cosmological constant including the RSD measurement of the cosmological growth rate, $f\sigma_8(z)$. The RSD measurement of growth data has been used to test cosmological models \cite{ref:RSDxu}. And a lower growth rate from RSD than expected from {\it Planck} was also pointed out in Ref. \cite{ref:RSDlower}.       

This paper is structed as follows. In Sec. \ref{sec:normal}, we give a brief review of the normal branch of DGP gravity and their modifications to the linear perturbation theory. We present the results of our MCMC study in Sec. \ref{sec:results}. Sec. \ref{sec:conclusion} is the conclusion. 

\section{Normal Branch of DGP gravity with a Cosmological Constant} \label{sec:normal}

In DGP model \cite{ref:DGP}, our Universe is a four-dimensional brane embedded in a five-dimensional Minkowski space-time described by the action
\begin{equation}
S=-\frac{1}{2\mu^2}\int d^5x \sqrt{-\hat{g}}\hat{R}-\frac{1}{2\kappa^2}\int d^4x \sqrt{-g} R+\int d^4x\sqrt{-g}\mathcal{L},
\end{equation} 
where the quantities with hats denote the five-dimensional ones; the Lagrangian $\mathcal{L}$ confined on the brane includes matter fields and brane tension or a cosmological constant; the constants $\mu^2=1/M^3_5$ and $\kappa^2=1/M^2_4$ are related to the reduced Planck mass in the bulk and brane respectively. The crossover scale $r_c=\mu^2/2\kappa^2$ governs the transition from five-dimensional to four-dimensional scalar-tensor gravity. 

By variation of the action, one obtains the modified Friedmann equation on the brane in a homogeneous and isotropic metric
\begin{equation}
H^2-\frac{\sigma}{r_c}\sqrt{H^2+\frac{K}{a^2}}=\frac{\kappa^2}{3}\Sigma_{i}\rho_i+\frac{\Lambda}{3}-\frac{K}{a^2},\label{eq:FErho}
\end{equation}
where $\sigma=\pm 1$ denotes the branch of the cosmological solutions; $\sigma=+1$ is for the self-accelerating branch and $\sigma=-1$ is for the normal branch; $H=\dot{a}/a$ is the Hubble parameter; $K$ is the spatial curvature; $a$ is the scale factor and $\rho_i$ are the energy components on the brane which include baryons $\rho_{b}$, cold dark matter $\rho_c$ and radiation $\rho_r$; $\Lambda$ is a cosmological constant or brane tension on the brane. The late time accelerated expansion is determined by two model parameters, the cosmological constant and the crossover scale. Using the definition of the dimensionless density parameter $\Omega_i=\kappa^2\rho_i(a=1)/3H^2_0$, the Friedmann Eq. (\ref{eq:FErho}) can be recast into 
\begin{equation}
E^2(a)=\left(\sqrt{\frac{\Omega_{m}}{a^3}+\frac{\Omega_{r}}{a^{4}}+\Omega_{\Lambda}+\Omega_{r_{c}}}+\sigma\sqrt{\Omega_{r_{c}}}\right)^2+\frac{\Omega_K}{a^2},\label{eq:FEomega}
\end{equation} 
where $\Omega_K=-K/H^2_0$, $\Omega_{\Lambda}=\Lambda/3H^2_0$, $\Omega_{m}=\Omega_{b}+\Omega_{c}$ and $E^2(a)=H^2(a)/H^2_0$ is the dimensionless expansion rate. Here $\Omega_{r_{c}}=(4H^2_0r^2_c)^{-1}$ satisfies the relation
\begin{equation}
\sqrt{\Omega_{r_{c}}}=\sigma\frac{1-\Omega_m-\Omega_{r}-\Omega_{\Lambda}-\Omega_{K}}{2\sqrt{1-\Omega_K}}.
\end{equation} 
The free model parameter can range in theory from zero to infinity. But noticing the fact that if $r_c\rightarrow \infty$ the general relativity is recovered and that the deviation from general relativity on solar system scale is also controlled by $r_c$, $r_c$ should have a very large value, at least be larger than the solar system scale. Actually, the cosmic observational constraint requires that $r_c>3H^{-1}_0$ \cite{ref:DGPPPF}. Therefore, to cover the limit case of general relativity, we take $\Omega_{r_{c}}$ is a free model parameter instead of $r_{c}$. It implies that general relativity is recovered when $\Omega_{r_{c}}$ is zero. As shown in Eq. (\ref{eq:FEomega}), the five-dimensional gravity effect becomes maximal when $r_c$ approaches to $H^{-1}_0$.  

To understand the properties of this modification to gravity, the Friedmann Eq. (\ref{eq:FErho}) can be rewritten in the following form \cite{ref:DGPconstraint1}
\begin{equation}
H^2=\frac{\kappa^2}{3}\Sigma_i\rho_i+\frac{\kappa^2}{3}\rho_{eff}-\frac{K}{a^2},
\end{equation}  
by the definition of an effective energy density of dark energy
\begin{eqnarray}
\rho_{eff}&=&\frac{1}{\kappa^2}\left(\Lambda+\sigma\frac{3}{r_{c}}\sqrt{H^2+\frac{K}{a^2}}\right)\nonumber\\
&=&\frac{3H^2_0}{\kappa^2}\left(\Omega_\Lambda+2\sigma\sqrt{\Omega_{r_c}}\sqrt{E^2(a)-\Omega_{K}a^{-2}}\right).
\end{eqnarray}  
It is clearly seen that the five-dimensional effect makes the effective energy density larger and smaller in the self-accelerating and normal branch respectively. From the continuity equation of the effective dark energy $\dot{\rho}_{eff}+3H(1+w_{eff})\rho_{eff}=0$, one can derive its equation of state
\begin{eqnarray}
w_{eff}=-1&+&\frac{1}{3}\frac{\sigma\sqrt{\Omega_{r_c}}\left(3\Omega_m a^{-3}+4\Omega_r a^{-4}\right)}{\Omega_{\Lambda}+2\sigma\sqrt{\Omega_{r_c}}\left[E^2(a)-\Omega_Ka^{-2}\right]^{1/2}} \nonumber\\
&\times&\frac{1}{\left[E^2(a)-\Omega_Ka^{-2}\right]^{1/2}-\sigma\sqrt{\Omega_{r_c}}}.
\end{eqnarray}  

In the longitudinal gauge, the scalar linear perturbations of the metric is given as
\begin{equation}
ds^2=-(1+2\Psi)dt^2+a^2(1+2\Phi)\delta_{ij}dx^i dx^j.
\end{equation} 
To derive the evolution of density and metric perturbations on the brane, the bulk metric equations are required \cite{ref:DGPconstraint3} where the quasi-static solutions with curvature was obtained. The Poisson equation and the traceless part of Einstein equations are given \cite{ref:DGPconstraint3} 
\begin{eqnarray}
k^2\Phi&=&\frac{\kappa^2}{2}\left[1-\frac{1}{3\beta}\right]\rho\Delta a^2,\\
k^2\Psi&=&-\frac{\kappa^2}{2}\left[1+\frac{1}{3\beta}\right]\rho\Delta a^2,
\end{eqnarray}
where $\beta$ is
\begin{equation}
\beta=1-2\sigma H^2 r_c\left(H^2-\frac{\Omega_K}{a^2}\right)^{-1/2}\left[1+\frac{\dot{H}}{3H^2}-\frac{2}{3}\frac{\Omega_{K}}{a^2H^2}\right],
\end{equation}
and $\Delta \equiv \delta+3aHv/k$ is the gauge-invariant comoving density contrast. Here $\delta=\delta\rho/\rho$ is the density contrast. The matter perturbation in the subhorizon ($k\gg aH$) satisfies the following equation \cite{ref:DGPstructure}
\begin{equation}
\Delta^{''}_m+\left[2+\frac{H'}{H}\right]\Delta'_m-\frac{3}{2}\frac{E_m}{E^2}\left[1+\frac{1}{3\beta}\right]\Delta_m=0, \label{eq:evolution}
\end{equation}  
$E_m=\Omega_{m}a^{-3}$. Here the prime $^{'}$ denotes the derivative with respect to $\ln a$. Via the definition of the gauge-invariant growth factor $f=d\ln\Delta/d\ln a$, which is reduced to $f=d\ln\delta/d\ln a$ in the matter comoving gauge, the perturbation evolution equation (\ref{eq:evolution}) can be rewritten as
\begin{equation}
f'+f^2+\left[2+\frac{H'}{H}\right]=\frac{3}{2}\left[1+\frac{1}{3\beta}\right]\Omega_m(a),\label{eq:f}
\end{equation}
where $\Omega_{m}(a)=H^2_0\Omega_{m}a^{-3}/H^2$ is the dimensionless dark matter energy density. Therefore, a tight constrain would be obtained by using the information from $f\sigma_8(z)$ which is almost model-independent and provides good test to dark energy models even without the knowledge of the bias or $\sigma_8$ \cite{ref:Song}. Recently, the authors of Ref. \cite{ref:fsigma8total-Samushia2013} compiled the redshift space distortion (RSD) based $f\sigma_8$ measurement, which amounts to eight $f\sigma_8$ data points in the redshift range $z\in [0.17,0.78]$. Another data point at $z=0.80$ was also obtained in Ref. \cite{ref:fsigma86-Torre2013}. For convenience, the data points are summarized in Table \ref{tab:fsigma8data}. The $f\sigma_8(z)$ data points have been used to constrain dark energy model; see Ref. \cite{ref:RSDxu} for a example.
\begin{center}
\begin{table}[tbh]
\begin{tabular}{cccl}
\hline\hline 
$\sharp$ & z & $f\sigma_8(z)$ & Survey and Refs \\ \hline
$1$ & $0.067$ & $0.42\pm0.06$ & 6dFGRS~(2012) \cite{ref:fsigma85-Reid2012}\\
$2$ & $0.17$ & $0.51\pm0.06$ & 2dFGRS~(2004) \cite{ref:fsigma81-Percival2004}\\
$3$ & $0.22$ & $0.42\pm0.07$ & WiggleZ~(2011) \cite{ref:fsigma82-Blake2011}\\
$4$ & $0.25$ & $0.39\pm0.05$ & SDSS~LRG~(2011) \cite{ref:fsigma83-Samushia2012}\\
$5$ & $0.37$ & $0.43\pm0.04$ & SDSS~LRG~(2011) \cite{ref:fsigma83-Samushia2012}\\
$6$ & $0.41$ & $0.45\pm0.04$ & WiggleZ~(2011) \cite{ref:fsigma82-Blake2011}\\
$7$ & $0.57$ & $0.43\pm0.03$ & BOSS~CMASS~(2012) \cite{ref:fsigma84-Reid2012}\\
$8$ & $0.60$ & $0.43\pm0.04$ & WiggleZ~(2011) \cite{ref:fsigma82-Blake2011}\\
$9$ & $0.78$ & $0.38\pm0.04$ & WiggleZ~(2011) \cite{ref:fsigma82-Blake2011}\\
$10$ & $0.80$ & $0.47\pm0.08$ & VIPERS~(2013) \cite{ref:fsigma86-Torre2013}\\
\hline\hline
\end{tabular}
\caption{The data points of $f\sigma_8(z)$ measured from RSD with the survey references.}
\label{tab:fsigma8data}
\end{table}
\end{center}

\section{Constraints on the Model} \label{sec:results}

To fix the background evolution history, the SNLS3 as standard candles \cite{ref:SNLS3} and BAO as standard rulers are employed. The initial conditions of the perturbations are fixed by the full information of CMB data which includes the recently released {\it Planck} data sets, which include the high-l TT likelihood ({\it CAMSpec}) up to a maximum multipole number of $l_{max}=2500$ from $l=50$; the low-l TT likelihood ({\it lowl}) up to $l=49$; and the low-l TE, EE, BB likelihoods up to $l=32$ from WMAP9; the data sets are available on line \cite{ref:Planckdata}. In the following part of this paper, the CMB data stets are denoted by Planck+WP. The following seven-dimensional parameter space is adopted 
\begin{equation}
P\equiv\{\omega_{b},\omega_c, \Theta_{S},\tau, \Omega_{r_c}, n_{s},\log[10^{10}A_{s}]\},
\end{equation}   
the priors for the model parameters are summarized in Table \ref{tab:results}. Furthermore, the Hubble constant $H_{0}=73.8\pm2.4\text{kms}^{-1}\text{Mpc}^{-1}$ \cite{ref:hubble} is adopted. The pivot scale of the initial scalar power spectrum $k_{s0}=0.05\text{Mpc}^{-1}$ is used in this paper.

In our analysis, we perform a global fitting to determine the cosmological parameters using the Markov Chain Monte Carlo (MCMC) method. The MCMC method is based on the publicly available {\bf CosmoMC} package \cite{ref:MCMC}. To compute the CMB power spectrum, we use the Boltzmann solver {\bf MGCAMB} \cite{ref:MGCAMB} which is a modified version of the {\bf CAMB} \cite{ref:CAMB} to include the modification of gravity. To use the RSD measurements of the growth rate $f\sigma_8(z)$, we added one module to obtain the corresponding likelihood. In this paper, instead of solving the differential equation (\ref{eq:f}), we derived the growth factor $f=d\ln\delta/d\ln a$ directly from the evolution of the perturbations at linear scales. Eight chains were run on the {\it Computing Cluster for Cosmos}, they were stop when the Gelman \& Rubin $R-1$ parameter $R-1 \sim 0.01$ was arrived; that guarantees the accurate confidence limits. 

\begin{figure}[htbp]
\includegraphics[width=9.5cm]{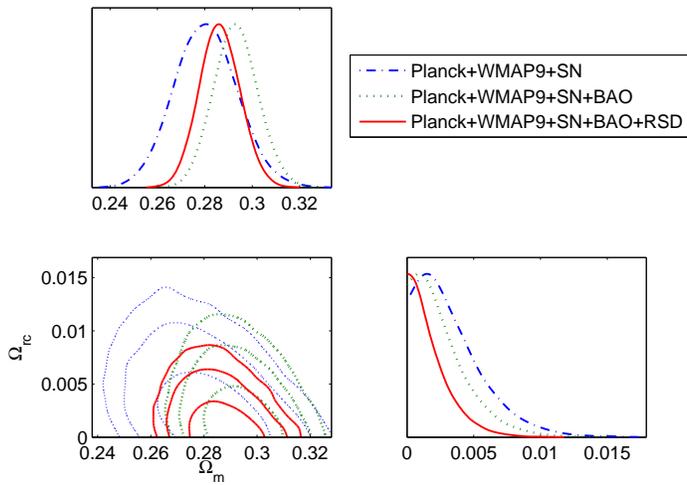}
\caption{The one-dimensional marginalized distribution on individual parameters and two-dimensional contours  with $68\%$ C.L., $95\%$ C.L. for nDGP model by using different combinations of SNLS3, {\it Planck}+WMAP9, BAO and RSD $f\sigma_8$ data sets.}\label{fig:nDGPcontours}
\end{figure}

At first, we take the SNLS3 plus Planck+WP as the basic data set for their basic abilities to fix the background evolution at low and high redshifts. With the addition of BAO and RSD one by one, the model parameter space was also scanned. The obtained results are shown in Table \ref{tab:results}. The one-dimensional marginalized distribution and two-dimensional contours for the model parameters $\Omega_{m}$ and $\Omega_{r_c}$ with $68\%$ C.L., $95\%$ C.L. and $99\%$ C.L. are shown in Figure \ref{fig:nDGPcontours}, where different combinations of cosmic observations are used.

\begin{widetext}
\begin{center}
\begin{table}
\begin{tabular}{cc@{}|cc@{}|cc@{}|cc}
\hline\hline  
Parameters & Priors & \multicolumn{2}{c|}{SN+Planck+WMAP9} & \multicolumn{2}{c|}{SN+Planck+WMAP9+BAO} & \multicolumn{2}{c}{SN+Planck+WMAP9+BAO+RSD}\\ \hline
$\Omega_b h^2$ & $[0.005,0.1]$ & $0.0222_{-0.00027}^{+0.00027}$ & $0.0224$ & $0.0221_{-0.00025}^{+0.00025}$ & $0.0222$ & $0.0222_{-0.00024}^{+0.00024}$ & $0.0222$\\
$\Omega_c h^2$ & $[0.01,0.99]$ & $0.117_{-0.0023}^{+0.0022}$ & $0.117$ & $0.119_{-0.0018}^{+0.0018}$ & $0.119$ & $0.117_{-0.0016}^{+0.0016}$ & $0.116$\\
$100\theta_{MC}$ & $[0.5,10]$ & $1.0416_{- 0.00060}^{+0.00060}$ & $1.0422$ & $1.0413_{-0.00058}^{+0.00057}$ & $1.0414$ & $1.0415_{-0.00055}^{+0.00056}$ & $1.0412$ \\
$\tau$ & $[0.01,0.8]$ & $0.0905_{-0.0145}^{+0.0129}$ & $0.0995$ & $0.0877_{-0.0137}^{+0.0121}$ & $0.0884$ & $0.0795_{-0.0126}^{+0.0115}$ & $0.0856$\\
$\Omega_{rc}$ & $[0,1]$ & $0.00335_{-0.00335}^{+0.000685}$ & $0.000842$ & $0.00264_{-0.00264}^{+0.000501}$ & $0.00121$ & $0.00183_{-0.00183}^{+0.000338}$ & $0.000458$\\
$n_s$ & $[0.5,1.5]$ & $0.967_{-0.00678}^{+0.00669}$ & $0.970$ & $0.962_{-0.005740}^{+0.00574}$ & $0.963$ & $0.965_{-0.00557}^{+0.00562}$ & $0.968$\\
${\rm{ln}}(10^{10} A_s)$ & $[2.4,4]$ & $3.0835_{-0.0277}^{+0.0253}$ & $3.101$ & $3.0825_{-0.0266}^{+0.0239}$ & $3.0821$ & $3.0616_{-0.0224}^{+0.0225}$ & $3.0698$\\
\hline
$H_0$ & $73.8\pm 2.4$  & $70.68_{-1.22}^{+1.24}$ & $70.20$ & $69.54_{-0.83}^{+0.84}$ & $69.24$ & $69.96_{-0.84}^{+0.78}$ & $69.71$\\
$\Omega_\Lambda$ & $...$ & $0.827_{-0.0558}^{+0.0469}$ & $0.774$ & $0.801_{-0.0500}^{+0.0400}$ & $0.775$ & $0.791_{-0.0468}^{+0.0341}$ & $0.757$\\
$\Omega_m$ & $...$  & $0.280_{-0.0129}^{+0.0129}$ & $0.284$ & $0.293_{-0.00917}^{+0.00913}$ & $0.295$ & $0.286_{-0.00832}^{+0.00839}$ & $0.285$\\
$z_{re}$ & $...$  & $10.971_{-1.117}^{+1.1345}$ & $11.696$ & $10.822_{-1.0848}^{+1.0786}$ & $10.887$ & $10.0251_{-1.0299}^{+1.0271}$ & $10.561$\\
$Y_P$ & $...$  & $0.245_{-0.000119}^{+0.000110}$ & $0.245$ & $0.245_{-0.000100}^{+0.000102}$ & $0.245$ & $0.245_{-0.000106}^{+0.000100}$ & $0.245$\\
${\rm{Age}}/{\rm{Gyr}}$ & $...$  & $13.70_{-0.050}^{+0.050}$ & $13.70$ & $13.74_{-0.040}^{+0.040}$ & $13.75$ & $13.73_{-0.038}^{+0.038}$ & $13.75$\\
\hline
\multicolumn{2}{c|}{$-\ln L$} & \multicolumn{2}{c|}{$5116.677$} & \multicolumn{2}{c|}{$5117.466$} & \multicolumn{2}{c}{$5124.063$}\\
\hline\hline
\end{tabular}
\caption{The mean values with $1\sigma$ error regions and the best fit values of model parameters and derived parameters for nDGP model with different combinations of SNLS3, BAO, {\it Planck}+WMAP9 and RSD data sets.}\label{tab:results}
\end{table}
\end{center}
\end{widetext}

\begin{center}
\begin{figure}[htb]
\includegraphics[width=9.5cm]{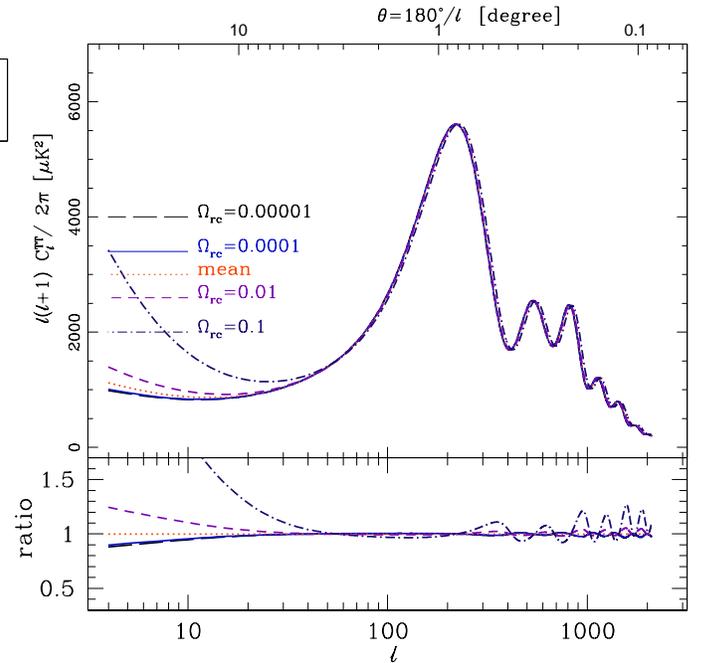}
\caption{The effects of model parameters $\Omega_{rc}$ to the CMB TT power spectrum for its different values, where the other relevant model parameters are fixed to their mean values as shown in Table \ref{tab:results}.}\label{fig:cls}
\end{figure}
\end{center}

As are shown in Table \ref{tab:results} and Figure \ref{fig:nDGPcontours}, the high quality of Planck+WP plus SNLS3 have constrained the model parameter space well by comparing to the previous results \cite{ref:DGPPPF}. Although the main constraint to the model parameter $\Omega_{r_c}$ for CMB data sets comes form the ISW effects at the large scales as shown in Figure \ref{fig:cls}, where the other relevant model parameters are fixed to their mean values as shown in Table \ref{tab:results}. It also clues that the determination of the background evolution, through the energy components in our universe, is also important to pin down the model parameter space. This point will be confirmed by the addition of BAO data points. It is widely believed that BAO data can give tight constraint to the background evolution through the standard ruler. The addition of BAO data set improved the constraint to the background relevant model parameters, such as $\Omega_{c} h^2$ and $\Omega_{r_c}$. And it is helpful to break the degeneracy between the model parameters $\Omega_{m}$ and $\Omega_{r_c}$ as shown by the green-dotted contours in Figure \ref{fig:nDGPcontours}. The model parameter $\Omega_{r_c}$ will modify the growth of matter perturbations through the Eq. (\ref{eq:f}), then a tighter constraint is expected when the growth relevant data sets are included. The results show that it is indeed. With all of the data sets, the mostly tight constraint $\Omega_{r_c}=1/(4H^2_0r^2_{c})=0.00183_{-0.00183}^{+0.000338}$ was obtained in this work. And our result suggest that $r_c$ should be around $11.688H^{-1}_0$. It implies that the five-dimensional gravity effect affect our Universe much weakly in the observable scale $H^{-1}_0$. As a comparison, the $\Lambda$CDM model was also constrained by using the same data combinations. The obtained $-\ln L$ for the $\Lambda$CDM model is $5123.771$. The difference between nDGP and $\Lambda$CDM model is $-\Delta\ln L=0.292$ for one extra model parameter. It implies that the nDGP model can fit currently available cosmic observations well as that for the $\Lambda$CDM model.  

\begin{figure}[htbp]
\includegraphics[width=9cm]{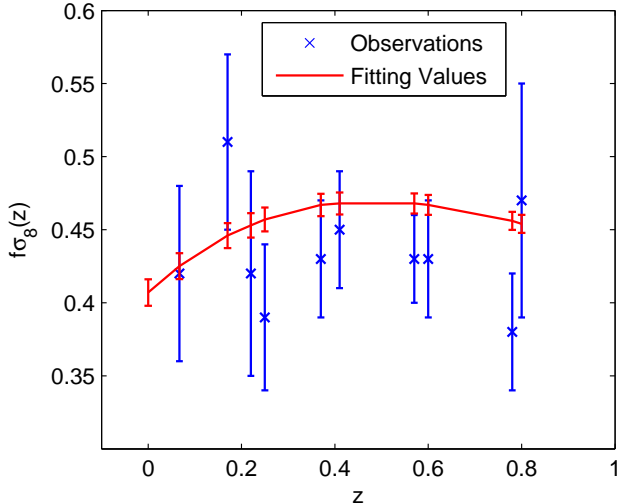}
\caption{The evolution of $f\sigma_8(z)$ with respect to the redshift $z$, where the 'Observations' denotes the observed values with $1\sigma$ error bars (the blue line segments with $\times$) as listed in Table \ref{tab:fsigma8data}, the central line (the red solid line) with error bars is the fitting mean values as listed in Table \ref{tab:results}.}\label{fig:fsigma8}
\end{figure}

We also showed the evolution of $f\sigma_8(z)$ (the central red solid line) with respect to the redshift $z$ with $1\sigma$ error bars in Figure \ref{fig:fsigma8}, where the observed values with $1\sigma$ error bars as listed in Table \ref{tab:fsigma8data} were also included (the blue line segments with $\times$). Here the values of $f\sigma_8(z)$ as derived model parameters are gathered in Table \ref{tab:fasigma8}, where the SNLS3, BAO, {\it Planck}+WMAP9 and RSD data sets are used.

\begin{center}
\begin{table}[tbhp]
\begin{tabular}{cccc}
\hline\hline 
$f\sigma_8(z=0)$ & $0.41_{-0.0090}^{+0.0090}$ & $0.41$\\
$f\sigma_8(z=0.067)$ & $0.43_{-0.0089}^{+0.0089}$ & $0.42$\\
$f\sigma_8(z=0.17)$ & $0.45_{-0.0086}^{+0.0085}$ & $0.44$\\
$f\sigma_8(z=0.22)$ & $0.45_{-0.0083}^{+0.0083}$ & $0.451$\\
$f\sigma_8(z=0.25)$ & $0.46_{-0.0082}^{+0.0082}$ & $0.46$\\
$f\sigma_8(z=0.37)$ & $0.47_{-0.0077}^{+0.0076}$ & $0.47$\\
$f\sigma_8(z=0.41)$ & $0.47_{-0.0075}^{+0.0074}$ & $0.47$\\
$f\sigma_8(z=0.57)$ & $0.47_{-0.0067}^{+0.0068}$ & $0.47$\\
$f\sigma_8(z=0.60)$ & $0.47_{-0.0068}^{+0.0067}$ & $0.47$\\
$f\sigma_8(z=0.78)$ & $0.46_{-0.0062}^{+0.0062}$ & $0.45$\\
$f\sigma_8(z=0.80)$ & $0.45_{-0.0061}^{+0.0061}$ & $0.45$\\
\hline\hline
\end{tabular}
\caption{The mean values with $1\sigma$ error regions and the best fit values of $f\sigma_8(z)$ at different redshift $z$, where SNLS3, BAO, {\it Planck}+WMAP9 and RSD data sets are used. In this case, the $\sigma_8$ is $0.82_{-0.012}^{+0.012}$ at redshfit $z=0$.}\label{tab:fasigma8}
\end{table}
\end{center}

\section{Conclusion} \label{sec:conclusion}

In summary, we have studied the normal branch of DGP braneworld gravity in the spatially flat case from the geometrical and dynamical perspectives through Markov chain Monte Carlo method.

On the geometrical side, the SN, BAO and CMB data from the first 15.5 months released {\it Planck} results are used to fix the background evolution. On the dynamical side, the growth rate $f\sigma(z)$ from the measurements of the redshift space distortion is used to determine the evolution of matter perturbations. The constrained results are summarized in Table \ref{tab:results}. As a comparison to the previous result \cite{ref:DGPPPF}, a tight constraint was obtained in this work. Our result suggests that the crossover scale $r_c$ should be around $12H^{-1}_0$ which is consistent with the previous result $r_c>3H^{-1}_0$ and greater. It also implies that the five-dimensional gravity effect is weak to be observed in $H^{-1}_0$ scale. As a comparison to the $\Lambda$CDM model, we found the difference between nDGP and $\Lambda$CDM model is $-\Delta\ln L=0.292$ for one extra model parameter. It implies that the nDGP model can fit currently available cosmic observations well as that for the $\Lambda$CDM model.  

\acknowledgements{The author thanks an anonymous referee for helpful improvement of this paper. L. Xu's work is supported in part by NSFC under the Grants No. 11275035 and "the Fundamental Research Funds for the Central Universities" under the Grants No. DUT13LK01.}

\end{document}